\documentclass[aps,prper,reprint, longbibliography, showkeys, nofootinbib]{revtex4-2}

\usepackage[T1]{fontenc}	
\usepackage{geometry}		
\usepackage{graphicx}
\usepackage{epsfig}
\usepackage[above,below]{placeins}	
\usepackage{times}
\usepackage{csquotes} 
\usepackage{gensymb} 

\usepackage{enumitem}          
\setlist{nosep} 

\usepackage{hyperref}  
\hypersetup{colorlinks=true,urlcolor=blue,citecolor=blue,linkcolor=blue}  

\usepackage{endnotes}

\usepackage[dvipsnames]{xcolor}

\usepackage{tabularx}
\usepackage{array}

\begin{document}
\title{Lessons from pendulums: A design comparison of three lab activities}

\author{Ian Descamps$^{1}$, Roger Tobin$^{1}$, Paul Wagoner$^{1}$, David Hammer$^{1,2}$, N. G. Holmes$^{3}$, and Rachel E. Scherr$^{4}$} 
\email[Please address correspondence to ]{ian.descamps@tufts.com}
\affiliation{$^1$Department of Physics \& Astronomy, Tufts University,
574 Boston Avenue, Suite 304, Medford, Massachusetts 02155, USA\\
$^2$Department of Education, Tufts University, 12 Upper Campus Road, Medford, Massachusetts 02155, USA\\
$^3$Laboratory of Atomic and Solid State Physics, Cornell University, Ithaca, New York 14853, USA \\ $^4$Physical Sciences Division, School of STEM, University of Washington Bothell,
Bothell, Washington 98011, USA}

\begin{abstract}
\label{abstract}
We present three versions of a pendulum lab activity to explore how theoretical commitments, motivations, and aspirations reflect in curriculum design.
In earlier work, Boudreaux \& Elby \cite{boudreaux_how_2020} discussed how different theoretical perspectives led them to different designs of tutorials. 
Here, we discuss our finding that even starting from the same theoretical perspective and having similar goals for students can lead to differences of design. 
We give three interacting reasons why our labs diverge: We have different expectations of students at our respective institutions; we have different ancillary goals; and we draw different implications from our shared theoretical commitments. 
Our account demonstrates the complexity of the relationship between theory, goals, and curriculum design. 
In this way, it adds to prior arguments for the importance of designers' articulating the reasoning for their choices as well as for the possible value of instructors' responsive adaptations to curricula.

\end{abstract}

\maketitle

\section{Introduction}
\label{intro}
For longer than any of the authors of this paper have been learning physics, introductory labs have included a pendulum activity. 
Often, a goal of the activity is for students to demonstrate that the period of a pendulum has a small but measurable dependence on the amplitude of its swing, a result at odds with the simple harmonic oscillator model and Galileo’s claim in Two New Sciences \cite{TwoNewSciences}. 
Such a lab provides a context for learning practices and values of empirical science: to see the dependence, students need both to take care in their measurements and to be aware of the precision they have obtained.\footnote{They do not need sophisticated technology! Mechanical stopwatches work fine.}

Each of us has assigned pendulum labs at the start of our introductory courses, at our respective institutions, with goals of introducing practices and values of empirical science. We want the activities to promote students' attention to experimental error as well as reflection and conversation about how science produces and assesses knowledge.
These goals reflect our shared view of students: that they have mostly experienced step-by-step instructions for experiments that demonstrate or validate theoretical claims from lecture, and need support to engage in more deeply scientific practices.

The similarities in our interests, goals, and views motivated research on how students may shift their approach to labs, to be less about “doing school” and more about “doing physics” \cite{smith_surprise_2018, smith_how_2020, smith_lets_2020}. That is, we have asked: when and how do students come to frame introductory labs in ways that support progress toward practices and values of empirical science? 
We were supported by the National Science Foundation for three years to collect and study data from our courses.\footnote{NSF DUE Awards: 2000394, 2000739, 2000711.} 
We paid particular attention to the pendulum labs, because they were the students' first engagement with activities different from their expectations. 
The project led to several publications analyzing student framings, including with respect to confirmation bias and epistemic agency \cite{sundstrom_problematizing_2020, phillips_not_2021, jeon_2023, sundstrom_instructing_2023, descamps_dynamics_2024}.

Throughout, we authors were aware of differences in our lab designs. We present the task differently: as testing the simple harmonic oscillator model $T=2 \pi \sqrt{\frac{L}{\textrm g}}$ (Cornell), as testing Galileo's claims (Tufts), or simply as studying what affects the period (UWB). 
We guide it differently, providing instructions for students' experimentation and statistical analyses (UWB and Cornell) or leaving almost everything up to the students (Tufts). 
The project was funded to study students' thinking, but as it ended we became interested to study our own: How did we come to design our labs so differently, given the similarities in our perspectives? 

\subsection{Theoretical frameworks}
We were inspired by Boudreaux \& Elby \cite{boudreaux_how_2020}, who described how their theoretical perspectives shaped their designs of tutorials on static friction. 
They demonstrated multiple ways in which their distinct theoretical orientations gave rise to differences in tutorials covering the same content.
Boudreaux worked from a “difficulties framework,” which attributes “a degree of stability in student conceptions”: “The strategy of challenging and de-settling non-normative ideas relies on those ideas maintaining coherence long enough to be held up, examined, and modified by the learner as they progress through the curriculum.” 
This perspective supported designing a tutorial to “elicit, confront, and resolve” incorrect conceptions. 

Elby worked from a "resources framework,” according to which “students’ responses to physics questions do not always correspond to stable conceptions.” 
Rather, patterns in students’ thinking depend on the context, “with contextual cues tipping students” to different ways of thinking. 
This perspective supported designing a tutorial to manipulate contextual cues that could tip students into activating different patterns of resources and guide them to notice and work to reconcile the differences. 

The authors were clear that explicit employment of theoretical frameworks did not fully determine their tutorial designs. They also drew on their intuitions and experiences, which they note are deeply entangled with their own theoretical dispositions: “theoretical orientations and instructional intuitions are not disjoint streams of knowledge that independently inform curriculum writing, but rather that they interact as they develop over time.”

\subsection{Our theoretical orientations}
In contrast to Boudreaux \& Elby, we see ourselves as starting from the same theoretical perspective. Like Elby, we all take a resource-based view of student knowledge, reasoning, and framing \cite{hammer_resources_2005}. 
The theoretical idea of “resources” is of “pieces” \cite{disessa_1993} of knowledge that can be active or not, depending on the context.
Some resources bear on physical mechanism, such as a basic sense that a stronger cause will produce a stronger effect, or that motion “dies away” if there is nothing to sustain it \cite{disessa_1993}. 
Some bear on kinds of activities and situations, influencing which features people recognize and pay attention to, as well as their judgments about appropriate actions. These resources are often referred to as “frames” and their activation as  "framing" \cite{tannen1993framing, hammer_resources_2005, scherr_student_2009}.

A resources perspective does not presume stability.
Resources may become active or not depending on the context. For example, in the context of cold weather, it is easy and intuitive to think “gloves keep hands warm,” while in the context of baking, “gloves keep hands cool.” Much of the work of physics is in connecting reasoning in contexts that start out as separate. Similarly, for example, someone framing what is taking place as following instructions may not notice or think to check whether the instructions make sense to them, as they would if they framed the moment instead as their participating in a decision.

Taking this perspective, we expect the dynamics of student thinking to be significantly influenced by how tasks are presented (e.g., \cite{Goodhew_resources_2021, Ford_game_2005, Frank_interactional_2012}), 
instructor or TA interactions (e.g., \cite{sundstrom_instructing_2023, wu_instructor_2022, Scherr_tainteractions}), intragroup social dynamics (e.g., \cite{conlin_making_2018, zhang_navigating_2021, scherr_student_2009, watkins_positioning_2018}), and practices established in the classroom community (e.g., \cite{appleby_disciplinary_2021, gouvea_motivating_2022}). 
In our prior work, we have considered dynamics at multiple scales of time and system, from individual minds in specific moments, to small collaborative groups over semesters, to classroom and school communities (e.g., \cite{scherr_student_2009, Frank_interactional_2012, radoff_its_2019, dini_case_2017, holmes_toolboxes_2017, quinn_group_2020}).

Our work in instructional labs has documented how students can frame lab as a matter of “jumping through hoops” \cite{phillips_not_2021}, and/or expect that success means confirming a known result \cite{smith_how_2020, descamps_dynamics_2024}.
These are laboratory versions of “doing school” \cite{jimenez-aleixandre_doing_2000}:  following “cookbook” instructions to arrive at the intended results, the findings expert authorities have determined to be correct. 
Unfortunately, these expectations are often at odds with the practices and values of empirical science.

The frames highlighted above, and students' framing in general, have profound consequences on the nature and quality of students' participation in labs. 
That is to say, how students frame the lab shapes the ways and extent to which they take up \textit{epistemic agency}. \cite{sundstrom_instructing_2023, descamps_dynamics_2024, phillips_not_2021, kalender_restructuring_2021}
Epistemic agency is the prerogative for individuals or groups to participate in “the whole range of components of knowledge building goals, strategies, resources, evaluation of results, and so on” \cite{scardamalia_higher_1991} (p. 108).
In our research and in our designs, epistemic agency is central to our understanding of what constitutes productive participation in scientific knowledge construction.

\subsection{Similarities and differences across our labs}
In designing our labs, we all anticipate that many students will arrive with expectations they have formed over years of experience in traditional labs, disposed to frame experimental work as a matter of following instructions to arrive at the correct (predetermined) conclusions \cite{smith_how_2020}. 
Part of what we all hope to accomplish with the pendulum lab is to disrupt that framing and to promote students' seeing lab as a place for genuine inquiry, such as to make sense of a surprising phenomenon \cite{smith_surprise_2018}. 
We expect students have resources for the beginnings of empirical science. Given the chance, students can and will be curious about the world, interested and able to try things to learn more about phenomena they notice. 

The instructional materials for the Cornell curriculum can be found on PhysPort; instructional materials for the UWB curriculum can be found on QUBES CourseSource  \cite{QUBES}; instructional materials for the Tufts curriculum are attached as supplemental materials.
As would be expected given our similar views and goals, our labs are similar in several ways. 
We all see the main purpose of introductory labs as helping students develop abilities in empirical science, rather than reinforcing concepts from lecture \cite{holmes_value_2017, smith_best_2021}.
The instructions, assessment, and feedback we provide emphasize practices of empirical science and challenge expectations that success means arriving at a “correct” answer. 
They give students time and autonomy to design, test, and refine their experimental methods and analyses, and all offer them significant latitude to draw and defend their own conclusions. 
Finally, all of our designs depend to some extent on “responsive teaching” \cite{ResponsiveTeaching}, that is the lab instructors' close attention to and engagement with student thinking, in individual and small group interactions, as well as in classroom community-level idea sharing and discussions.

Still, as we noted above and will elaborate below, our labs differ in several significant ways. 
We decided to look into those differences, and in what follows we report on what we have found: Variations in our designs arise from (1) differences in our expectations of students at our respective institutions, (2) variations in our ancillary pedagogical goals, and (3) nuances of our theoretical perspectives. 
We will try to make our reasoning transparent, to expose uncertainties and flexibilities in our designs, to argue for expectations of complexity, and to motivate views of instruction and instructional design as ongoing and dynamic. 

We hope that our comparisons provide insight into and concrete examples of how learning theories guide curriculum development. Scholars have long argued, and empirical evidence indicates, that it is important for teachers implementing curriculum to understand the purposes and aims of its design \cite{shulman_those_1986, shulman_knowledge_1987, ball_reform_1996, davis_designing_2005, harlow_structures_2010, elby_rethinking_2020, robertson_curricular_2021}. 
Our comparative analysis supports these arguments, exposing "hidden decisions" \cite{harlow_structures_2010} and "pedagogical judgments" \cite{davis_designing_2005} that led us to different versions of the pendulum lab.

In the next section, we review the literature on the design of “non-traditional” instructional labs, with an emphasis on their theoretical bases, both implicit and explicit. 
Then we move on to our curricula:  first, in sec. \ref{contexts}, we discuss each of the institutional and instructional contexts; in sec. \ref{Pedagogical Overviews}, we describe our respective approaches and their ongoing evolution in turn and in some detail; and in sec. \ref{Comparisons} we compare these approaches. 
In sec. \ref{Discussion} we step back to make sense and discuss the implications of our comparison before, finally, concluding in sec. \ref{Conclusion}.

\section{Literature Review:  Lab Curriculum Research and Development}
\label{Lit Review}
Physicists, educators, and education researchers have long expressed dissatisfaction with traditional laboratory instruction \cite{otero_past_2017, smith_best_2021, reif_teaching_1979, thornton_tools_1987,may_historical_2023}.
We understand “traditional” labs to be defined by \textit{i)} aims to reinforce content knowledge, \textit{ii)} activities that demonstrate or verify physical concepts, \textbf{and} \textit{iii)} prescriptive experimental protocols.  
Despite such labs' goal of supporting conceptual learning, evidence indicates they are generally unsuccessful at it \cite{thornton_tools_1987, etkina_design_2010, holmes_toolboxes_2017, smith_direct_2020}.
Research also indicates that traditional labs negatively impact the development of expertlike epistemological attitudes and beliefs \cite{wilcox_impact_2016, wilcox_open-ended_2016, wilcox_students_2017, walsh_skills-focused_2022}.

Reshaping laboratory instruction with an eye towards authentic disciplinary and experimental practices has notable historical precedent \cite{otero_past_2017, may_historical_2023}.
A number of research projects have developed and documented alternative curricula, often inspired by educational theory: 
\begin{itemize}
    \item RealTime Physics (RTP; \cite{sokoloff_realtime_2007, thornton_tools_1987, thornton_learning_1990}) 
    \item Investigative Science Learning Environment (ISLE; \cite{etkina_design_2010, brookes_implementing_2020}); 
    \item Thinking Critically in Physics labs, first at the University of British Columbia \cite{holmes_doing_2013, holmes_making_2014, holmes_impact_2015}
    and continuing at Cornell \cite{holmes_operationalizing_2019, kalender_restructuring_2021, phillips_not_2021, sundstrom_instructing_2023};
     \item the instructional labs at CU-Boulder (e.g., \cite{lewandowski_initial_2018, werth_impacts_2022}) 
    \item the Design, Analysis, Tools, and Apprenticeship lab (DATA lab; \cite{funkhouser_design_2019}).
\end{itemize} 

Some of these learning environments have documented evidence that they support productive experimental activity in both quantity and quality \cite{etkina_design_2010, holmes_developing_2020} and improve students’ epistemological attitudes and beliefs \cite{wilcox_open-ended_2016, smith_direct_2020, brewe_modeling_2009}. 
At the same time, generalizing the motivations of these curricula as engagement in genuine experimental inquiry or skills-focused designs loses the specificity necessary for researchers to rigorously evaluate and compare them and, crucially, for educators to effectively adapt and implement them.
These lab curricula vary in their specific goals, theoretical influences, and pedagogical features and structures.
In the next section, we discuss these curricula in more detail by focusing on the differing roles of theory in their development and documentation.

\subsection{Theoretical influences on lab curricula}


\subsubsection{Semi-Empirical}
The DATA and CU-Boulder lab curricula are presented as the conclusion of a largely empirical transformation process. 
In each case, the lab developers interviewed interested faculty about their goals for laboratory education, synthesized the themes of those interviews into consensus goals, and then iteratively refined the goals along with faculty members.
These consensus goals provided the basis for the design of the curriculum.
We refer to this process as semi-empirical because, while the overall course transformations incorporated empirical research, the design decisions were not explicitly grounded in empirical (or theoretical) explanations.

For the DATA lab, they settled on four learning goals: experimental process, data analysis, collaboration, and communication. Notably, the “design was divorced from the specific physics content.” This feature was explicitly motivated by their empirical findings rather than by theoretical considerations: “the learning goals developed from a faculty consensus design did not include specific content”  \cite{funkhouser_design_2019}.
To select and design instructional materials and course structures, Funkhouser \textit{et al.} applied Constructive Alignment \cite{biggs_alignment_1996}:  the alignment of an instructional system based on constructivist principles. 
In this scheme, course objectives should articulate how students can demonstrate understanding through actions; students should be put in situations that can elicit such “performances of understanding,” and students should be assessed based on their performance.

The DATA lab is a stand-alone course and two semester sequence. 
The typical lab activity takes place over two weeks and students generally follow the same two-week trajectory: pre-lab homework informs the qualitative exploration and prediction phase, and the first week culminates in students’ designing their experimental procedure; in the second week, students conduct their investigation and iteratively improve their procedure as they analyze their data and generate interpretable results.

Details of the initial CU-Boulder lab curriculum, also a stand-alone lab course, can be found in a series of PERC papers \cite{lewandowski_student_2018, pollard_impact_2018, pollard_transforming_2018, lewandowski_initial_2018}.
The design of this curriculum was motivated by five main learning goals:
\begin{itemize}
    \item students’ epistemological attitudes should align with experts, 
    \item students should have positive attitudes about the course,
    \item students should have positive attitudes about experimental science,
    \item students should be able to generate a graph showing model and data, and 
    \item students should demonstrate “set-like” reasoning about data.
\end{itemize}
Additionally, in other work \cite{werth_impacts_2022}, the CU-Boulder team refers to their initial lab as seeking to develop skills. The research team also notes that, due to various constraints, the course retained many features of a traditional lab \cite{pollard_impact_2020}.

Nevertheless, Pollard \textit{et al.} \cite{pollard_impact_2020}  highlight several transformed features they expect to be salient for student learning: 
for a pre-lab assignment, students watch short, interactive videos;
the experimental activities require students to measure a previously unknown quantity or outcome, as opposed to verifying values learned in lecture or listed in a textbook; 
students use these measurements to make a prediction related to their experiment and then, once they measure that prediction, compare results with other groups. 
That comparison is expected to support the intentional use of uncertainty values. 
Additionally, most of the lecture sessions (which are infrequent) are dedicated to measurement uncertainty concepts.

Although presented as a mostly empirical process, we contend that there is non-trivial theoretical work involved in the development and design of these curricula.
For example, a feature of the DATA lab is that, for a given activity, each lab group conducts a different experiment, with no set end point.
Funkhouser \textit{et al.} mention that this feature “has added benefits for [each] student's ownership and agency of the work.”
It could be that ownership and agency were understood by the designers to support productive engagement and thus were part of the decision-making process. Another possibility is that increased ownership and agency were serendipitous benefits afforded by this choice, which was influenced by other considerations. 

This approach obscures the justification of design decisions.
Curricular features are instead presented as naturally emerging from learning goals.
As a result, the documentation of curriculum transformations does not provide actionable insight for educators enacting the curriculum or selecting among alternatives as they adapt design features.
The work of specifying how to accomplish learning goals can be productively understood as a theory-laden process of making hypotheses or conjectures about how to best support learning \cite{sandoval2014,Cobb2007}. 
Accounts of the theory-laden reasoning that motivates design decisions in a curriculum are especially important given our understanding of the variation in how educators take up research-based instructional methods \cite{dancy_how_2016, turpen_construction_2010, affriyenni_navigating}, the consequential role of the broader (activity) system within which specific design features are implemented \cite{tasar_characterizing_2023, harlow_learning_2020}, and the ways in which an awareness of this reasoning can support responsive instructional moves \cite{robertson_curricular_2021}

\subsubsection{Theoretically Informed}
The curricula described in this section all explicitly invoke learning theories (and related empirical research) as they describe their designs, albeit in different ways: 
the design principles of RTP were inspired by educational research;
the researchers at CU Boulder outline expectations about how authenticity supports learning and list various features of their CURE expected to promote it; 
the designers of ISLE list their overarching instructional values and intentionalities that help them select theoretical frameworks to do design work. 

As one of the first evidence-based or research-based reformations of laboratory instruction, the RTP lab curriculum is a significant scholarly and historical milestone for PER as a field \cite{may_historical_2023, otero_past_2017, thornton_tools_1987}.
In addition to  learning goals, Sokoloff \textit{et al.} \cite{sokoloff_realtime_2007} delineate design principles “based on education research,” that undergird the RTP labs. 
The lab activities
\begin{itemize}
    \item are sequenced to provide students with a coherent observational basis for understanding a single topic area in one semester or quarter of laboratory sessions;
    \item provide activities that invite students to construct physical models based on observations and experiments;
    \item help students modify their common conceptions about physical phenomena that make it difficult for them to understand powerful general principles of physics;
    \item work well when performed in collaborative groups of two to four students;
    \item incorporate Microcomputer-based laboratory (MBL) tools so that students can test predictions by collecting and graphing data in real time;
    \item incorporate a learning cycle consisting of prediction, observation, comparison, analysis, and quantitative experimentation;
    \item provide opportunities for class discussion of student ideas and findings; and
    \item integrate homework assignments designed to reinforce critical concepts and skills.
\end{itemize}

The central innovation of the RTP labs are the MBL tools: sensors and interfaces that enable the real-time collection, display, and analysis of various physical measurements.
These MBL tools were designed “with the student learner in mind” \cite{thornton_tools_1987}.
They enable students to focus on the physical world, giving them the ability to explore and learn from the world; the immediate feedback of the devices extends the range of potential investigations, and the graphical display options can provide students a start in developing representational competence \cite{thornton_learning_1990}.
The design of the MBL tools, along with the RTP lab curriculum, builds off of findings from cognitive science and education research at the time (e.g., \cite{mcdermott_student_1987, halloun_common_1985, arons_achieving_1983}) as well as the 1998 AAPT laboratory recommendations \cite{american_association_of_physics_teachers_goals_1998}. 


ISLE similarly aims for students both to learn concepts and to engage in empirical inquiry, but differs from RTP in how it implements experimentation. 
ISLE aims more explicitly at supporting students' development in practices of empirical science. 
As part of that, it shifts responsibility for experimental design onto students, trading richer experience for the efficiency and reliability of successful, canonical findings. 
Like RTP, ISLE involves “cycles” of learning and experimentation, but the form differs. 
RTP emphasizes student predictions and hypotheses, in part to elicit and respond to student misconceptions. 
ISLE, in contrast, begins with observational experiments and “something that needs explaining” \cite{brookes_implementing_2020}, in part for the designers' views of student epistemologies as essential features of their engagement and targets for instruction. 

To describe the theoretical foundations of their curriculum, Brookes \textit{et al.} \cite{brookes_implementing_2020} overview their “core intentionalities” \cite{Garrison_MacMillan} – their beliefs and values about what their teaching is trying to accomplish – and use them as criteria for evaluating potential theoretical frameworks \cite{Cobb2007}.
With these criteria in mind, they discuss the “bricolage” of theoretical perspectives that actualize their beliefs about how learning happens and how to support it. 
This discussion expands on their 2010 paper \cite{etkina_design_2010}, which highlighted \textit{interpretive knowing} as a desired form of cognition and activity that guided their design.
Such theoretical descriptions provide not only specificity around important design decisions but also context for how and to what end the theories were implemented. 

In response to the need to teach remotely during the beginning of the COVID-19 pandemic, CU-Boulder redesigned their lab course to be a Course-based Undergraduate Research Experience (CURE) \cite{werth_impacts_2022}. 
The authentic discovery that defines CUREs made it an attractive option for their redesign, as it supports engagement in key scientific practices, including collaboration and iteration; furthermore, the emphasis on authenticity (with respect to professional physics research) is expected to support motivation and persistence. 
In other words, \textit{authenticity} as a theoretical construct shaped many of the design decisions in the construction of this curriculum.

The designers expected the authenticity of this research to support learning in two main ways: 
first, authentic research requires scientific practices such as reading published literature, developing a research plan, data analysis, collaboration, and iteration; 
second, the authenticity of this research goes \textit{beyond} “offer[ing] students glimpses of what it means to do experimental physics” \cite{smith_best_2021}, which is expected to positively impact student motivation and affect \cite{oliver_student_2023}. 
Oliver \textit{et al.} \cite{oliver_student_2023} identify six different aspects of the course designed to engage students in such authentic research:  
a scaffolded literature review, meetings with the PI of the project (twice per semester for each group), metacognitive reflections, teamwork, peer review, a final reflection structured as a “memo to future researchers,” and the publication of their research. 

The education research from the first few semesters of this course highlights that engagement in scientific practices did occur \cite{oliver_student_2023}. Students felt they participated in authentic research \cite{werth_impacts_2022, oliver_student_2023}, that they had achieved their teamwork goals \cite{werth_assessing_2022}, and that teamwork contributed to their success in the course \cite{werth_assessing_2022}.


As with learning goals, theory does not uniquely and completely determine design \cite{Cobb2007, elby_rethinking_2020, boudreaux_how_2020}. 
Still, theoretically-grounded descriptions provide insight into the design process and the reasoning that links learning goals and pedagogical structures. 
Empirical study is necessary to understand what works when, for whom, and how, but the application of those findings to new situations and curricula necessarily involves theorizing.
Explicating the reasoning for design decisions, especially theoretically-grounded reasoning, can generate instructionally actionable insight \cite{elby_rethinking_2020, robertson_curricular_2021} and empirically testable conjectures \cite{disessa_cobb_2004,Cobb2007, sandoval2014}.

\section{Instructional Contexts}
\label{contexts}
Although throughout this paper we refer to the authors' theoretical commitments, we should be clear that these labs were designed and are implemented collaboratively.
At Cornell, Emily Smith, Phil Krasicky, Mark Lory-Moran, Michael Niemack, Jared Maxson, Cristina Schlesier, and Rebeckah Fussell all contributed to the curriculum design, in addition to co-author  Holmes. 
Holmes and her colleagues were supported by an internal grant to do the research and development for \textit{Thinking Critically in Physics}.\footnote{\url{https://www.physport.org/curricula/thinkingcritically/}}
At Tufts, Timothy Atherton and Hugh Gallagher contributed to the curriculum design, in addition to co-authors Hammer, Tobin, and Wagoner. 

Cornell is a large, private, selective research university; Tufts is a small, private, selective research university; and UWB is one of two small campuses of a large, public research university.
UWB runs on the quarter system, and the lab course is part of a 3 term sequence; both Cornell and Tufts are on the semester system.

The introductory lab at Cornell is a stand-alone course, and the lab curriculum at UWB is an adaptation of the Cornell curriculum. Each includes a lecture, where students reflect on their labs and work through activities to develop statistical skills. 
At Tufts and UWB, introductory labs are sections connected to the main lecture courses. Tufts has no lab-specific lecture and lab instructions change in small ways from one semester to the next.
Here we focus on the spring instantiation of the Tufts labs, taught by Hammer, specifically from the spring 2024. In Fall semesters, the labs are shared across algebra-based and calculus-based courses, and the overall course management proceeds differently.

The labs at Cornell are instructed by graduate TAs, they last two hours each, and there are a maximum of 20 students per section. 
At UWB, professors instruct the labs, which last two hours and have a maximum of 24 students per section.
The Tufts labs are instructed by graduate and undergraduate TAs, last two and half hours, and there are a maximum of 14 students per section.

At both UWB and Tufts, the pendulum lab activity occurs over three lab sessions; at Cornell the activity takes two lab sessions.   
The lab courses at Cornell and UWB assign homework; Tufts labs do not (students complete all notes and  reports within the 2.5-hour sessions).

\section{Three Versions of the Pendulum Labs}
\label{Pedagogical Overviews}
Our designs share foundations, orientations, and aspirations. 
All take as central a desire to engage students in doing empirical science, focusing on “what it means to do experimental physics: the approach, techniques, skills and ways of thinking when conducting authentic physics experiments” (\cite{smith_best_2021}, p. 1).
Crucially, this doing must be active and intentional\textemdash  it must be epistemically agentive. 
We design our labs in such a way as to encourage and support students taking up roles as knowledge agents, who have the ability to creatively and intentionally pursue goals relevant to knowledge production. 

In the subsections that follow, we provide a more in-depth description of the pedagogical logic, an outline of the design itself along with an explanation of the \textit{reasoning} for the various design decisions. 
In these descriptions, we seek to make clear how differences arise in our designs from different prioritizations and emphases of that overarching goal as well as different “secondary” considerations.

\subsection{Cornell}
At Cornell, the design of the pendulum activity creates a context for experimental practices to have purpose and relevance by setting students up to encounter an empirical-theoretical discrepancy. 
The instructional materials for the Cornell pendulum lab can be accessed through PhysPort.\footnote{\url{https://www.physport.org/curricula/thinkingcritically/}}
The lab instructions discuss the pendulum period equation derived from a model (with assumptions listed), and the guidance of the in-lab instructions set students on a path to produce data that indicates a deviation from this model. 
Students are assessed on their process and the quality of their work, rather than by whether they arrive at the correct conclusion.

At the heart of this lab activity is the epistemological richness of evaluating a model in conversation with experimental work; 
the (potential) conflict creates a need not only for calculating and interpreting uncertainty, but also for experimental design and data visualization skills that work together to respond to the theoretical-experimental discrepancy. 
This rich context aligns with the overarching course goal of students developing a more sophisticated understanding of what it means to know things and to construct knowledge in physics. 

As the first lab activity, the structure of this lab aims to \textit{cultivate} student agency. 
The guidance of the in-lab instructions constrains the field of inquiry, but in ways that provide students with a sense of direction for their work so as to avoid the frustration of not knowing what to do. 
The instructions are open-ended to create decision-making opportunities. 
For example, in the first lab the experimentation begins with a direction to set up a pendulum and record period measurements at 10° and 20° but students are not told how to carry out this experimentation. In this way, the in-lab instructions begin to shift responsibility for experimentation and knowledge production onto the students. 

Before the experimentation begins in the first lab, the first several questions of the in-lab instructions prompt groups to discuss and reflect on how they are going to approach collaboration. 
Students are asked to discuss their purpose and goals in this course and how they will support each other’s goals; they are asked to synthesize a common or collective goal for their group and to generate a plan of some sort for ensuring fairness and equity in the division of labor. 
Their recorded responses to these questions make up the groups’ “partner agreement.” 
In addition to the ways this discussion gets groups to proactively think about how they are going to collaborate, the written record can serve as a useful artifact for navigating tension and disputes between group members.

After this discussion, there is one more thing students must do before they begin their experimentation: properly connect their CoLab or Jupyter Notebook to their utilities folder. 
The instructional team at Cornell recently updated their in-lab instructions to be written in Jupyter Notebook format, combining the in-lab instructions with students’ lab notebook. 
Students can use text cells to write out responses to questions or record other notes; the document also includes some base code that students can use to input and process their data. 
While coding and computational skills are not explicit foci of this course, including Jupyter Notebook/CoLab as tools in this course allows students to become familiar with them and see them as useful.

Once the code is properly set up, students begin their experimentation. 
Students start by constructing a pendulum and recording period measurements at 10° and 20°. 
They then perform some calculations (computing the mean, standard deviation, and standard uncertainty) and answer some questions connecting those statistical measures to the physical causes of variability and uncertainty in their data. 
This work launches them into the main focus of the first session: reducing uncertainty. 
Groups can choose between two options for their main experimental work of the session:  exploring how increasing swings per trial reduces uncertainty or
comparing the approaches of using repeated single swing trials to a multi-swing trial.

In between lab sessions, during the lecture portion of the course, students participate in a contrasting cases invention activity on quantifying distinguishability. 
The activity introduces students to the “t-prime” (t') statistic \cite{holmes_quantitative_2015}, similar to Student's t-test, with the idea that the contrasting cases/invention structure supports student's conceptual understanding of this statistical tool. 
In the following lab activity, students are prompted to use and interpret t' values for the data they produced in the first session. 
The goal of the contrasting cases invention activity is to provide students with a tool they understand (as opposed to a black box) that they can use to make meaning from their data.

At the beginning of the second lab session, students compute t' values for their data and are then asked about “reasonable next steps or new questions based on your set of t' values?” 
The main focus of the second lab session is pushing models to their limits, and they are expected to use t' to help them in that work.
As in the first session, groups can choose between two options; the two options this week are “reducing uncertainty” or “pushing on assumptions.” 
In both cases, students must explain how their task helps test the limits of the model and better understand the physical phenomenon. 
All groups generate ideas for possible experiments and then must “systematically test multiple measurement methods” for their chosen task. 

The instructions also prompt groups to consult and collaborate with other groups. 
First, when groups are generating ideas for possible experiments, the instructions suggest that students consult other groups. 
Then, once groups have conducted their systematic tests, the instructions direct students to find another group who chose the other task option and to exchange results with them. 
This exchange and collaborative discussion forms the basis for yet another round of iterating and improving their experimental design. 
A desired outcome of these activities, and of the first lab session as well, is that students use their ideas, data, and tools to make decisions about their experiment and to construct knowledge about the physical phenomenon. 

The directions and prompts to reduce uncertainty and test the limits of the model increase the likelihood that students encounter an empirical discrepancy. 
This discrepancy establishes a purpose for their experimental practices; in this context, ideally, students perceive a genuine need for careful consideration of uncertainty, statistical analysis, data visualization, and experimental design. 
At the same time, recognizing and responding to this discrepancy is not a requirement of the lab, and the assessment practices reflect that. 
The assessment emphasis is not on the content of their conclusions (whether they produce “the correct answer”) but rather on their process and explanation. 

Instructors expect that this activity will surface and cue students’ expectations to verify authoritative knowledge and follow prescribed instructions. 
This approach aims to elicit and then respond to, if not confront, misaligned or improper expectations.
At the end of the activity, through instructor feedback on their lab report and discussion in the associated lecture, students can reflect on their process and approach. 
Altogether, the lab activity aims to gradually transform students’ expectations about what instructional labs are about and cultivate their agency.
During this lab unit and throughout the course, the specificity of the instructions is scaled back so that students take up more responsibility for figuring out what to do. 

\subsection{UWB}
The lab curriculum at UWB is an adaptation of the \textit{Thinking Critically in Physics} curriculum described above. 
The instructional materials for the UWB pendulum lab can be accessed through QUBES CourseSource \cite{QUBES}.
The design of this pendulum lab emphasizes meaningful, intentional use of experimental practices and the importance of collaboration and community feedback. 
The overarching goal, of this activity and the lab course in general, can be summarized as supporting students to engage in intentional, purposeful activity and perceive their work as consequential for producing knowledge about physical phenomena. 
Moreover, science is done with people, in community, with established (but transformable) social and epistemic norms. Collaboration, within groups and among the whole class, is a central feature of the curriculum.

Lab groups are tasked with generating a “team agreement” in the first week, and revisiting and updating it in the second week. 
The presence and clarity of a teamwork agreement in their notes is an explicit dimension of the grading rubric for the first two weeks. 
In the first lab session, lab instructors are encouraged to highlight a handout that discusses various potential roles one \textit{could} take on during group work  (e.g., skeptic, theorist, principal investigator).
Students also use a collaborative lab notebook, where students color-code their contributions to the document as they progress through the in-lab instructions.

The experimentation begins with a single pendulum for the whole class, whose period students measure after it is released from 10° and 20°. 
Importantly, students are given no instructions on how to do so. 
They are also are asked to make a prediction about how the periods at 10° and 20° will compare, "based on the physics that you know."
Each individual measurement is recorded in a shared data table, and then the students (in their lab groups) are tasked with creating a histogram of these measurements to visualize the variability. 
This task also involves determining the precision of their measurement and calculating the mean and statistical uncertainty (standard deviation) of the data. 
This specific activity seeks to both direct student attention towards particular concepts  (the role of data visualization and the physical meaning of uncertainty) and to foster student agency. 

There is a tradeoff pertaining to students' agency here: it is not the students’ choice to construct a histogram. 
At the same time, constructing a histogram involves choice and deliberation.
Students have to figure out what an appropriate bin size is and what “appropriate” means in this context. 
They are responsible for making the measurements in the first place, and, once they construct a histogram, they are responsible for interpreting it. 
There is no predetermined correct answer their histogram is supposed to reveal; the purpose of the task is to do data interpretation and make sense of this phenomenon. 
Additionally, by directing students to create a histogram, they have a starting point or foothold in beginning to think about data visualization and the relationship between variation and uncertainty. 
The agentive tradeoff aims to forestall the inhibitive effects of discomfort or frustration with not knowing what to do \textbf{and} encourage choice and agency within the loose constraints of the activity.

The second week begins with a short activity on comparing data sets quantitatively, which first asks lab groups to invent a way to distinguish example measurements and then summarizes some key points using real-life examples connected to the interests of the instructor (e.g., when Scherr teaches lab, she discusses datasets around classroom achievement in different instructional conditions). 
The invention activity primes students to notice and attend to particular quantities when interpreting and comparing data. 
Furthermore, both as an in-class, small group discussion and as part of their homework, students are asked “What are two data sets for which knowing whether they are the same or not would make a difference to your life or your community?” 
This reflection question aims to highlight the utility and meaningfulness of the data interpretation practices they will be engaging in during the second (and third) lab sessions. 

Following the mini-lecture and invention activity, the bulk of the experimental activity for the second lab revolves around reducing uncertainty. 
The students are tasked with using their data and experience to iteratively refine their experimental approach. 
The in-lab instructions direct students’ attention and activity towards certain concepts, such as measuring consecutive periods and collaborating with other groups, while also requiring explanations and interpretations of this work. 

Additionally, the goal of reducing uncertainty provides a way to show students that \textit{their activity} is consequential and important in this learning environment:  
at the end of the first week, students are asked to brainstorm ideas about how to reduce experimental uncertainty in measuring pendulum periods and then in the second week they are told to pick one of those methods to implement. 
Many students do not seem to expect that their answers could inform their activity in the lab: they often list unhelpful options such as "eliminate human error" or "reduce air resistance."
The follow up in the second week provides another opportunity for their coming to see lab as a place where their ideas are consequential.

The third week begins with an introduction to the t' statistic \cite{holmes_quantitative_2015}, which the students apply to the data sets they produced in the previous week. 
Instructions prompt students to interpret their calculated t' values and make decisions based on those interpretations. This again constrains their activity, but they remain responsible for generating interpretations, decisions, and conclusions.
By introducing the t' statistic in this way and scaffolding its use, the lab activities provide students multiple opportunities to use the t' statistic to derive meaning from and interpret their data. 

Throughout, the lab instructions provide students with tools (e.g., a histogram template, a statistics vocabulary sheet, and the t' statistic) and then ask them to make meaning through their use. 
Even if the students do not initially recognize these tools or their use of them as meaningful, the repeated written prompts direction to improve their experiment work to shift them in that direction. 
During lab meetings, the ways in which instructors engage, question, and respond to students similarly work to reinforce this framing, that the students' use of these tool should have purpose and meaning.

After calculating t' and identifying ways to improve their experiment, students return to the overarching experimental question:
\begin{displayquote}
“At this point, you are done with structured questions and ready to take the reins of your own investigation. Pick one of your proposed experiments and work with your group to design and carry out your own high-precision experiment. In your design, identify the possible outcomes and generate multiple possible explanations for each outcome. The goal is to answer the question: \textbf{Does the period of your pendulum depend on the amplitude of the swing (for 10° vs 20°)?}”
\end{displayquote}

The experimental prompt is neutral with respect to outcome: It provides no theoretical context about simple harmonic motion or Galileo's claims that favor specific predictions.
Partway through the third lab session, the instructor sets up a shared data table for groups to fill in with their period measurements, uncertainty values, t' values, and conclusion. 
This provides a communal artifact that sparks inter-group discussion about results and conclusions. 

It is not until after they have turned in their lab notes for grading that students are introduced to the simple harmonic model and associated period equation. 
Homework questions ask them, first, to restate their conclusion about the distinguishability of periods at 10° and 20°, and second, whether their conclusion aligns with the model. They are then asked to reflect on how knowing the “correct” result of an experiment can create bias, with a specific prompt to generate strategies to mitigate it.

\subsection{Tufts}
At Tufts, the written instructions change in small ways from semester to semester; examples are attached as supplemental materials. They are designed to create an environment in which students recognize the need for experimental practices, in particular of attention to precision and of questioning authority. 
This fits an overarching goal for the course, including the "content" focused lecture portion, that students learn how to learn, including to assess the quality of their knowledge, to make decisions, and to take actions they see as reasonable.
This perspective views students, especially at the start of the semester, as poised to frame lab as following prescribed steps toward predetermined correct results, "doing school" \cite{jimenez-aleixandre_doing_2000} in ways most have experienced. 
The labs try to promote "doing science," to shift responsibility onto students for figuring out how to approach, organize, and carry out their experimentation \textbf{and} for generating knowledge claims from that experimentation.
To that end, the written materials are minimal, just half a page long for each of the first two sessions. They present the empirical challenge but provide no instructions on how to accomplish it.
 
The hope is that students feel empowered by the agency and responsibility afforded to them.
Part of that, instructors anticipate, is students' learning to manage feelings of uncertainty over what to do, without the clarity of steps to follow. Ideally, they come to enjoy the autonomy, including the challenges of not-knowing. 
That is, the labs are designed and taught in ways that may promote "meta-affective learning" \cite{jeon_2023, radoff_its_2019}). 
In labs as in the courses as a whole, instructors and teaching assistants try to recognize confusion as productive, responding to students favorably (e.g. "great!") when they express feeling unsure and prompting them to elaborate (e.g., "talk about what has you confused").

The first activity asks students: “\textit{How precisely can you measure the period of a pendulum?}” 
The instructions clarify the answer should be quantitative, "e.g. to within 0.2 seconds, or 0.1, 0.0000001, whatever you can manage." 
The TA shows students materials available; there is a rich assortment. From there, they are responsible for building their own pendulums and deciding both how to be precise and how to quantify that precision. 
The session includes time for students to share their techniques and results.

Throughout the building of the pendulum and data collection, the TAs ideally focus on listening and supporting the substance of student ideas. 
Often students turn to standard deviation to quantify uncertainty, based on their experience in labs, and, in these cases, TAs respond along the lines of “What does standard deviation mean to you?” or “What does it tell us about data?” 
Most students are unclear about what it means; they see it as a black box algorithm they are "supposed to" use, a part of framing lab as "doing school."  
The TA should discourage using algorithms in this way; students should approach labs in ways that make sense to them.

With limited time, and to avoid prompting students' expectations of following prescribed steps, there is no instruction or guidance toward established statistical measures of variation or representational forms. 
A tacit element of the reasoning for this choice, which this comparative analysis brought to light, is a belief that students have sufficient capabilities with statistics and data representation to generate sensible approaches.
If there were more time available, it would be possible to follow students' inventions with explicit instruction, as in \textit{Thinking Critically in Physics}.

The instructions for the second week present the task as follows:
\begin{displayquote}
“[During the previous lab session], we heard conjectures that have history to them: There were some groups saying the angle (or the “amplitude”) of the pendulum doesn’t affect the period; others said it does. Galileo Galilei (1564-1642) studied this and claimed the period of a pendulum does not depend on amplitude, or on the mass of the weight (called the ‘bob’). The challenge today is to study these questions: To the precision you are able to measure, does the amplitude (angle) affect the period? Does the mass of the bob?” 
\end{displayquote}
This is the spring 2024 version of the instruction sheet. 
Every year, the lab teaching team reads and considers possible revisions. Indeed, the first sentence above, about student conjectures, was written in response to discussions TAs reported having in their sections. 
This allows the instructions to be responsive to student ideas and builds up the agency of the TAs themselves \cite{harlow_learning_2020}.
A year-over-year consistent feature of this experimental prompt is the mention of Galileo’s claims. 

Semantically, the question is not confirmatory and the following paragraph begins “Please take care to be honest about your data and your findings!” 
(That sentence was added based on observations in previous semesters, of students setting their own evidence aside.)
Still, the inclusion of Galileo’s name in the instructional prompt typically leads students to expect to confirm his claims (the famous scientist must be correct!). 
That expectation sets up an empirical discrepancy.
In most sections, some lab groups will produce data that indicates (to them and/or to a knowing instructor) that period does change with period. 
This potential discrepancy generates a need for experimental practices; data visualization, uncertainty estimation, and an understanding of the \textit{production} of data have functional relevance to the genuine need to make sense of this discrepancy.\footnote{The expectation to confirm Galileo may also reflect or cue students' framing lab as doing school, but see \cite{descamps_dynamics_2024} for evidence it can support productive experimental practices.} 

In the third session, lab groups present and discuss their conclusions.\footnote{In 2026, as we prepare this manuscript, the lab team decided to keep the pendulum lab to two sessions.}
This discussion is termed a “meaning-making session,” where the whole class attempts to come to a consensus regarding the experimental question.
The instruction sheet for this session includes a translated quote from the 11th century Arab scientist and mathematician al-Hasan ibn al-Haytham, describing how seekers of truth should only submit to argument and demonstration and “make himself an enemy of all that he reads” \cite{Sabra_ibn}.
The instruction sheet explicitly says that the values of reasoning and empirical evidence, as opposed to deference to authority, are core to the course. 
In addition to presenting their work to their peers, students discuss and reflect on this quote in light of their work in the first two weeks.

Throughout the three-week activity, the instructional team emphasizes that they do not grade for the correctness of students' results but for the quality of their engagement in conducting their investigation. 
When the labs are complete, with comments and scores ready for students, the instructor brings up the topic in the main course lecture, first to affirm the independence with respect to mass (and to discuss how that coheres with Newtonian theory), and second to consider the variation with amplitude. 
Almost always, students’ data provides the basis for claiming there is a dependence, and that sparks a discussion about the role of authority in science.

\section{Comparisons}
\label{Comparisons}

We all begin the semester with a pendulum lab we design (and redesign) based on a resources-based view of student knowledge, reasoning, and framing, with shared objectives regarding epistemological framing: labs are where students can and should engage in constructing knowledge through empirical study of physical phenomena. 
We all emphasize to students that the goal is not “getting the right answer,” as they might expect, but rather learning to engage in high-quality, intellectually rigorous experimentation. But we design these labs in different ways. Table 1 lists the differences across our curricula that we identified in this work.

\begin{table*}
\begin{tabularx}{0.9\linewidth}{|X|| X |X |X|}
 \hline
 Curriculum Feature & Cornell & UWB & Tufts \\ [0.5ex] 
 \hline\hline
 Experimental question & Model-testing, asked at beginning of unit & Neutral phrasing, asked at beginning of unit  & Galileo-testing, asked in second session \\ [0.5ex] 
 \hline
 Statistics and representation guidance & Instructions prompt students to make and use histograms and t' & Instructions prompt students to make and use histograms and t' & No explicit written guidance \\ [0.5ex] 
 \hline
 Medium of materials & Instructions and lab notes in CoLab or Jupyter Notebook & Instructions in view-only documents, lab notes in jointly-edited documents & Instructions in PDFs, lab notes as shareable documents \\[0.5ex] 
 \hline
 Lab homework & Tutorials on statistics and uncertainty & Reflection questions & None \\[0.5ex] 
 \hline
 Group discussions & Dedicated lecture time and at the discretion of lab instructor  & Mini-lectures at the beginning of lab sessions and planned check-ins throughout & Third lab session is dedicated whole group discussion \\ [0.5ex] 
 \hline
 Collaboration support & Partner agreement and roles handout & Partner agreement and roles handout & None \\[0.5ex] 
 \hline
 Instructors & Graduate TAs  & Professors & Graduate and Undergraduate TAs \\[0.5ex] 
 \hline
 Course Logistics & 20 students per section, sessions last two hours & 24 students per section, sessions last two hours & 14 students per section, sessions last two and a half hours \\[0.5ex] 
 \hline
\end{tabularx}
\caption{Overview of differences among the three lab activities}
\end{table*}

In this section, we compare our approaches with respect to presentation, written guidance, and responsive teaching. 
We selected these aspects of the curricula to compare because they connect tightly to our goals of student agency and shared theoretical commitments regarding framing and agency. 
To make sense of the differences in our curricula, we focus on the \textit{design reasoning} involved in our decisions.

\subsection{Presentation of the task}
\label{EE}

At Cornell, the instructions present an equation for the period, described as part of the simple harmonic model, along with some explanation of what a model is. At Tufts, the instructions present Galileo's claims that the period does not depend on amplitude or mass. Additionally, at Cornell, the lab unit as a whole is introduced as a model-testing one from the outset while at Tufts it is only in the second (of three) sessions that Galileo’s claims are introduced, with reference to student ideas or findings from the first session. 
At UWB, the prompt is neutral, simply asking students to investigate whether the period of a pendulum depends on the amplitude.

Part of the reasoning for providing a model that students are meant to test, as at Cornell and Tufts, is to cue up expectations of confirmation.
For Hammer, the mention of Galileo sets up an eventual reflective discussion about the role of authority in science and the importance of evidence and argumentation.  
For Holmes, the similar approach with a mathematical model sets up an epistemological version of “productive failure” \cite{kapur_designing_2012}. 
The expected discrepancy between the provided model and students' data generates an epistemological rich context for underscoring the importance of empirical model-testing. 
One reason for the more granular in-lab instructions at Cornell is to guide students to find the discrepancy. 

An additional aspect of the reasoning for Hammer and Holmes is an expectation, borne out of their years of teaching these labs, that many students will have some awareness of the pendulum period equation and/or Galileo. In Scherr's experience, that expectation does not apply for students at UWB. 
She chooses a neutral phrasing so that the curriculum emphasizes empirical science as constructing facts in a community with shared standards. When disagreements about conclusions arise – as is a common occurrence in all our labs – the UWB curriculum still supports students in discussing notions of trust, accountability, and authority. Absent a reference to external authority, that discussion remains rooted in the students’ own data and experiences. 
 
We all agree on the important epistemological expectations we aim to communicate to students in the first lab. 
We all present the task, with an elicit-confront-resolve approach or neutral phrasing, as a way to support students' framing the lab as centered on their reasoning.
The differences reflect our perceptions of the student populations in our courses. They also reflect differences in personal preferences: model-testing is a central and specific idea for the designers at Cornell while at Tufts and UWB the focus is on constructing scientific knowledge and “science as a refinement of everyday thinking.” 

\subsection{Written guidance}
\label{in-lab}
One of the main differences among our curricula – indeed the initial inspiration for this paper – is in the written guidance we provide (or do not provide) for students. The lab instructions at Cornell and UWB specify particular tasks and activities; the instructions at Tufts do not. 

For Holmes and Scherr, their more detailed instructions address a concern that students may feel lost or overwhelmed to a degree that would inhibit their learning. 
The concern for Hammer, on the other hand, is that the length and specifications of instructions can cue students' “doing school” framing. 
If there are no instructions to parse, the possibility of the instructions cuing or stabilizing that framing is not possible.

Holmes and Scherr take the perspective that this common framing can be a potentially productive resource: use student expectations that doing lab means following instructions, and design those instructions carefully to direct them towards productive activity, such as to make a histogram. In subsequent sessions, instructions become sparser, giving students more responsibility for their decisions.

Having the students use t' to quantify distinguishability plays out differently at UWB and Cornell.
At UWB, the instructions guide students to consider the possibility of a subtle dependence. At Cornell, finding that dependence conflicts with the model they likely expected to confirm. At both institutions, it is up to the students to decide what to conclude. 
In this way, both curricula use students' willingness to “do school” as a way to scaffold productive activities, reducing the guidance over time (within the first lab activity and across the term as a whole) to cultivate agency.

Here, we see a difference in our theoretical stances towards student behavior and learning, within the same general framework. We differ in our expectations about the dynamics of students' framing, in particular over when and how they may come to frame lab as "doing science" for themselves. 
Can students' willingness to adhere to instructions \textemdash as "jumping through hoops" \cite{phillips_not_2021} or "doing school" \cite{jimenez-aleixandre_doing_2000}\textemdash support their progress toward productive scientific work, or is it something to disrupt from the start?
We also have different perspectives on the risks of students' feeling lost or frustrated: Scherr and Holmes view these emotions as a barrier to engagement and participation, whereas Hammer hopes that students grow from wrestling with such feelings, that the lab is an opportunity for meta-affective learning \cite{radoff_its_2019}. 
(We all, meanwhile, agree it is important that students feel safe to try things and explore, including to trust that their grades are not based on arriving at prescribed outcomes.)

Recent work in these labs has highlighted the complexity of the epistemic and affective dynamics of student engagement. 
Studying labs at Cornell, Sundstrom \textit{et al.} \cite{sundstrom_instructing_2023} observed that successful instructor interventions to shift students away from confirmation framing were not always associated with an increase in measures of student agency. 
And, studying labs at Tufts, Descamps \textit{et al.} \cite{descamps_dynamics_2024} observed students' confirmation framing was not always associated with a lack of agency; in fact, it seemed to drive productive practices of investigation.

Additionally, Jeon \textit{et al.} \cite{jeon_2024} highlight a student's complex affective experiences of confusion in the Tufts labs: she insisted that she “hates physics,” during side conversation with her lab mates, because their confusing experimental results defied her expectations that physics should align with common sense. At the same time, this students' feelings of confusion motivated her to take up more epistemic agency during the lab.
Thinking in terms of a tradeoff between giving students autonomy and managing their feelings of uncertainty might be a useful heuristic for design decisions, at least within non-traditional labs, 
Data from study of students' engagement, however, suggest the effects of different choices are not so clear-cut as “tradeoff” might imply.

As theoretical constructs, framing and agency provide insight into various kinds of design decisions, but they can be taken up in ways that lead to different outcomes. 
One might focus on decision-making agency \cite{holmes_developing_2020} or on framing as a phenomenon of an individual’s cognition, but these are far from the only ways to understand and apply framing and agency to pedagogy.
Additionally, we cannot ignore the influence of secondary and contextual factors in the divergence of our in-lab guidance structures.
At Cornell and UWB, that the instructions provide guidance allows students to do “mini-activities” aimed at building up collaboration skills. 
The minimal lab instructions at Tufts have the effect of shifting various aspects of classroom management onto the TAs, which is aligned with Hammer’s active interest in TA education, which we take up in the next section.

\subsection{Responsive Teaching}
\label{instructors}
We all encourage responsiveness, but, with minimal written guidance, the Tufts labs depend more critically on the instructors' practices of attending, interpreting, and responding to their students' thinking. 
In part, the differences in our expectations of instructors reflect differences in expectations for the materials. \textit{Thinking Critically in Physics Labs}, the core of the materials used at Cornell and UWB, is designed for wide dissemination and use. It is difficult in general to rely on instructor responsiveness, and so these materials provide more detailed in-lab instructions to guide students toward the intended, rich arenas of experimental inquiry: making useful plots, quantifying uncertainty, refining methods, generating conclusions, and sifting through empirical conflicts. 

In contrast, the Tufts labs are only for use at Tufts. That has the advantages and disadvantages of the instructors' freedom and responsibility to use their judgment as they work with students. We see this as a second “agentive tradeoff” that comes with the more detailed in-lab instructions, analogous to the tradeoff for students:  The instructors, too, are guided and constrained with respect to what they may notice and how they may respond. 

At UWB, Scherr expects the instructors, who are all faculty, to act with autonomy and agency, including noticing, assessing, and responding to student work during implementation. Scherr's own modifications to the \textit{Thinking Critically in Physics Labs} curriculum reflect this aspect of her context. For example, when Scherr introduces the activities around quantitatively comparing datasets, she uses examples from her personal and research interests, while other instructors use different examples. 
Still, the detailed in-lab instructions function to provide clear representation of pedagogical intentions, while at the same time taking a familiar form, and in these ways may facilitate instructors' adoption. 

At Cornell, the instructors are graduate TAs, who come in with a wide range of experience and interest in teaching, familiarity with research-based pedagogical approaches, beliefs about how learning does and should happen, and confidence in the classroom. Moreover, as with the faculty at UWB, there is limited room for their preparation or supervision, although student TAs are not free to modify the materials. The detailed in-lab instructions function to minimize variation across the TAs and to make it easier for them to implement the curriculum as intended. 

At Tufts in contrast, part of the objective for Hammer is TA education and preparation. The TAs are a mix of graduate students and undergraduates, the latter often coming to the position after successful experiences as Learning Assistants \cite {Otero_learningassistants} in the highly interactive lectures.  
The minimal written instructions have the effect of shifting responsibility to the TAs for noticing, interpreting, and responding to students. Often this involves helping the students to shift in their framing of lab, such as when students look to the TA for prescriptive guidance over how to proceed.  

Hammer's practice has been to meet with the team of lab TAs for several hours each Friday. 
The TAs recount snippets of what they noticed in their students' participation over the week, and they choose examples of students' written work. They compare interpretations, following practices of educator support for responsive teaching \cite{robertson_responsive}, and ideas for how they might respond. They also discuss plans for the coming week, including to suggest refinements to the coming assignment's wording and design. 
Experience at Tufts makes it clear that these meetings play an important role in ensuring TAs implement the curriculum as intended and develop skills in responsive teaching.

\section{Reflections and Discussion}
\label{Discussion}
As we discussed in the introduction, the purpose of our work here is to wade into the gap between our theoretical perspectives and our curriculum designs.
Like Boudreaux \& Elby \cite{boudreaux_how_2020}, the inspiration for our comparison, we compare our designs with the hope of elucidating how our theoretical orientations shape the curricula.
While they sought to connect the differences in their theoretical frameworks to the differences in their designs, we have sought to make sense of the differences in our design despite the alignment of our theoretical perspectives.

First, we review the similarities, many of which reflect our theoretical alignment. 
We all aspire to promote student epistemic agency within the process of experimentation and to center the substance of student ideas. 
We all work to make sure the experiments in our labs are not exercises in confirming or verifying concepts from lecture;
try to shift students away from problematic framing; expect instructors to be responsive to the substance of student ideas; build into our labs the generation and use of data representations; highlight the need for some way to quantify uncertainty, and, in general, aim at communicating to students that their ideas and their data are of central importance.

These similarities reflect a shared understanding of how framing can influence the agentive element of student behavior, of how, when given the chance, student can be capable of productive scientific work, and how a wide variety of contextual cues can tip students into different framings.
These similarities and the underlying theoretical connective tissue between them punctuates the motivation to consider how and why we arrive at designs with such substantive differences.

\subsection{Presentation of Activity}
Above, we highlight how each of our labs present the activity to students:  At UWB, the experimental question is neutral; at Cornell, the experiment is presented as a model-testing activity; at Tufts, students are asked to investigate the claims of a renowned authority.  
In the section \ref{EE}, we connected these differences to variations in the expectations we seek to cultivate in students.

The neutral phrasing of the UWB instructions clearly focuses the experimental work on the students. 
We expect the references to the simple harmonic model (Cornell) and Galileo (Tufts) to elicit potentially problematic expectations about verifying already-known, authoritative theoretical claims, so that we can confront and disrupt these expectations.
Moreover, the reference to these models sets students up to experience a disagreement between theory and experiment, which in turn generates a genuine (for the students) need to engage in experimental practices of working to reduce uncertainties, use and interpret statistical tools to distinguish data sets, and extend experiments to test emergent questions.

\subsection{Structure of Guidance}
The instructions at Cornell also have the effect of scaffolding students' collecting data that will indicate a theoretical-empirical discrepancy. 
Tufts takes a different, minimalist approach to written instructions. 
In sec. \ref{in-lab}, we highlight how our shared understandings of framing, agency, and affect inform our differing decisions about instructions.
To an extent, the difference in our approaches to in-lab instructions indicate a difference in theoretical perspective\textemdash or perhaps difference in how we understand the instructional implications of theory.

For Hammer, students' framing the activity as “doing school” \cite{jimenez-aleixandre_doing_2000} is a primary concern that drives many of the decisions affecting the structure of in-lab guidance.
For Scherr and Holmes, while this framing certainly presents a concern, they take the approach of using students' expectations to follow instructions as a way to nudge students towards productive arenas of experimental work and scientific reasoning.
That is, students themselves may not initially see the point of making a histogram and so do so because the instructions tell them to, but thinking about how to best make a histogram and comparing and interpreting that histogram is a rich context for reasoning about and with representations.

There is an additional, affective dimension to Scherr's and Holmes's decisions around in-lab guidance:  the detailed instructions seek to ensure that students do not feel lost or overwhelmed.
To be clear, while the instructions at Cornell and UWB are more detailed \textit{relative to Tufts}, they are much less detailed than a “cookbook” lab.
Indeed much of their design work here involves writing open-ended prompts that provide guidance to students while also shifting decision-making onto students.

Across the curricula, the designs reflect different choices over an “agentive tradeoff” in the level of detail to provide.
There is also an “affective tradeoff,” with respect to students' experiences of uncertainty, greater at Tufts with less written guidance. Students learning to manage and regulate potential feelings of discomfort\textemdash meta-affective learning \cite{radoff_its_2019} is itself a desired outcome.
While the in-lab instructions at Cornell and UWB provide support for students with respect to agency and affect, at Tufts that falls much more to the instructor, a graduate or undergraduate TA.
In this way, TAs at Tufts are responsible and empowered to be more flexible and adaptable in how they respond to specific students and emergent idiosyncrasies.

Of course, the lab instructors at Cornell and UWB are also empowered and able to respond to students, but the onus is reduced. Part of the reasoning for more granular instructions at Cornell and UWB is that they minimize the potential negative influence of inexperienced TAs or instructors with discordant pedagogical beliefs. Indeed, the Tufts lab design places a heavy demand on TAs, and experience has shown the importance of supporting their learning as instructors. This is another form of tradeoff: Maintaining that support in the face of varying enrollments (and thus numbers of TAs), levels of engagement and skill of both the TAs and the faculty assigned to that role, has proved to be challenging. There has been noticeable variation in the effectiveness of the labs across different semesters.

Finally, we note a key difference with respect to curricular objectives with respect to statistical analysis. At UWB and Cornell, much of the guidance within the materials concerns the definition and use of the t' statistic. 
At Tufts, there is no “content” objective in this lab with respect to canonical statistics. 
Rather, the objective is entirely epistemological, that students recognize and take on the challenge of quantifying uncertainty.
The difference in course structure (stand-alone lab course versus component of a lecture course) plays a role in this decision-making: the stand-alone structure affords opportunities to engage in invention activities about statistics.

\subsection{Models of Students}
Moving forward, one area in particular where this exercise in comparison has provided inspiration for each of us is the degree to which our design decisions are informed by often unstated “working models” of the students. 
At Cornell and Tufts, experience has taught Holmes and Hammer that students in their labs will have heard of both Galileo and the simple harmonic period equation, while that is not Scherr’s experience at UWB. These considerations were consequential in how we operationalized our theoretical commitments.
Scherr and Holmes both have a sense of the affective consequences for their students of a lack of in-lab instructions: in part, the guiding nature of their in-lab instructions is designed to meet students where they are. 

We refer to these expectations as working models to reflect both their unfinished and partially tacit nature, and to highlight these expectations as connected to our theoretical commitments. 
Implicit in our desire to shift students’ framings is an expectation of how students will think and act coming into our labs. 
Research has shaped our expectations \cite{hu_examining_2018,hu_examining_2018-1,hu_qualitative_2017,Hu_Zwickl_2017,smith_how_2020,smith_lets_2020,smith_surprise_2018,sundstrom_problematizing_2020,descamps_dynamics_2024,scherr_student_2009}, along with our experiences in and out of the classroom.
Still, as we engaged in our comparisons, we often found ourselves needing to explicate and clarify our understandings of the students. 
In reflecting on the ways in which the Cornell and UWB labs structure student engagement with statistical and representational tools, the Tufts instructional team realized the lab design was shaped by unstated expectations about students’ abilities to generate and use novel representational forms. This motivated a redesign of the third lab session to include more explicit attention to data representation and analysis.  

Additionally, in discussing our various expectations of students – as individuals – we also recognized how structuring collaborative groups remains an open question. 
Collaboration is built into our curriculum on a fundamental level, and at each school, we engage in a handful of strategies to promote the emergence of equitable and supportive small groups. 
Still, we are all uncertain how to best to support groups' working effectively, as well as how to assess "effectiveness" of collaboration.

\section{Closing thoughts}
\label{Conclusion}
The differences among our lab activities demonstrate that a theoretical framework of epistemic agency and framing can engender multiple, distinct forms of pedagogical structures.Our labs provide a sample, but they certainly do not span the entire design space.
There is plenty of opportunity, for curriculum designers as for instructors, to innovate, to explore, to try things out, as well as to conduct scholarship that deepens our understanding of what works. 
Given the heterogeneity of national and global lab contexts \cite{Holmes_landscape2020, Geschwind_global2024}, there is not only opportunity but also necessity, for local inquiry by instructors, as well as for the development of curricula that are flexible and adaptable to local circumstances.

In reflecting on how curriculum developers and the broader PER ecosystem can work towards more useful research outputs, Elby \& Yerdelen-Damar introduced the notion of \textit{instructionally generative fodder}: descriptions of curriculum, related research of student activity and learning in designed contexts, and more general research that provides actionable insight into learning and how to support it \cite{elby_rethinking_2020}. 
To optimize the generativity of curricular products, they suggest that developers make the rationale for design decisions available and clear to instructors, including points of uncertainty. 

In recent years, there has been a push for PER scholars to include more detailed and more useful descriptions of theory in curriculum development. In doing this work, researchers have commonly focused on design decisions and choices\textemdash making more decisions explicit and connecting them to theoretical frameworks. 
We suggest a slight shift in focus, to instead emphasize \textit{design reasoning}. That is, descriptions of curriculum should bring to the fore the reasoning involved in how design decisions are made and how a curriculum is expected to mediate learning. This suggestion is largely a rephrasing and synthesis of Elby \& Yerdelen-Damar’s \cite{elby_rethinking_2020} discussion of generativity in curricular products;
it also echoes other arguments that it is important for teachers to understand the purposes and aims of curricular materials \cite{shulman_those_1986, shulman_knowledge_1987, ball_reform_1996, davis_designing_2005, harlow_structures_2010, robertson_curricular_2021}.

Consider how the experimental question is posed in our three pendulum lab activities. Focusing on design decisions, one might say that our understanding of student agency and framing guided our decisions about how to phrase the experimental question. This explains the decision, but provides no insight into how we arrived at different phrasings. Focusing on the design reasoning requires us to discuss expectations of students (at Cornell and Tufts) to be aware of the pendulum period equation, to describe how these questions lead into an epistemologically productive failure and elicit-confront-resolve approach, and to articulate how each of these phrasings work to emphasize the substance of students' data and experimental work.

A shift to design reasoning also echoes the argument that Brookes \textit{et al.} \cite{brookes_implementing_2020} outline in their paper on the motivations for ISLE. 
Building from educational theorists MacMillan \& Garrison \cite{Garrison_MacMillan}, they “suggest that if we are to make better connections between education theory and classroom implementation, both researchers and implementers need to articulate their underlying intentionalities more explicitly.”

Robertson \textit{et al.} \cite{robertson_curricular_2021} illustrate the utility of knowledge of what we are calling design reasoning in a study of Learning Assistants in a university-level physics course. They specifically provide evidence that “knowledge of the purposes of questions or sequences of questions within [research-based instructional materials] [...] can support (or serve as a resource in) the enactment of responsive instruction” (p. 151). 
Robertson \textit{et al.} build from a long tradition of education scholars who maintain that making visible the "pedagogical judgments" \cite{davis_designing_2005} and "hidden decisions" \cite{harlow_structures_2010} of curricular materials can support teachers in enacting theoretically oriented and research-based curricula \cite{shulman_those_1986, shulman_knowledge_1987, ball_reform_1996, remillard_examining_2005}.
We extend this claim to recommend that curricular materials and research on curriculum development in PER ought to foreground design reasoning.

Our examination of our own curricula helped us recognize and articulate aspects of the reasoning that play significant roles in our developing and refining activities and materials.
We contend that a focus on design reasoning can help make descriptions of curricula more generative for practitioners. Design reasoning can also be generative for researchers: expectations about how a curriculum will support learning should inform studies of what takes place. 
Given that research-based pedagogies are taken up in noticeably different ways (if they are taken up at all) \cite{turpen_construction_2010, dancy_how_2016, affriyenni_navigating}, it would be useful for research to highlight the consequential aspects of curricula and crucial interactions.

\acknowledgments{We would like to thank the faculty and staff at Cornell, UWB, and Tufts involved in developing and implementing these labs. We acknowledge the Cornell University College of Arts and Sciences Active Learning Initiative for providing support for part of the development of the \textit{Thinking Critically in Physics} curriculum. We also thanks members of the Tufts STEM Education program, in particular Sophia Jeon, Miguel Vasquez-Vega, Ira Caspari, and the students in the Qualitative Research in STEM Education course for feedback on initial comparative analyses. Finally, we appreciate the thorough feedback of our anonymous reviewers, in particular the connections to Robertson \textit{et al.} \cite{robertson_curricular_2021}
}

\bibliography{references}
\end{document}